# Diffusion Limited Growth in Systems with Continuous Symmetry.

Umberto Marini Bettolo Marconi
*Dipartimento di Matematica e Fisica, Università di Camerino, Via Madonna delle Carceri,I-62032 , Camerino, Italy*

Andrea Crisanti
*Dipartimento di Fisica, Università di Roma "La Sapienza" P.le A.Moro 2, I-00185 , Roma, Italy*

To study the effect of slow heat conduction during phase separation, we discuss the relaxation properties of an $O(N)$ symmetric model with Phase field type dynamics, where a non conserved order parameter field couples bilinearly to a diffusive field. In the limit $N \to \infty$ we obtain an exact solution. The analysis reveals three different types of growth regimes and a very rich dynamical behavior. Finally the connection with the Mullins-Sekerka instability is expounded.

Many important phenomena occurring in nature can be described as growth processes. For such a reason these are attracting an increasing attention from theorists and experimentalists alike. It is well known [1–3] that simple models can lead to very rich dynamical behaviours such as steady-state patterns, growing structures, moving fronts, dynamical scaling, etc.. A great deal of study has been dedicated to the relaxation dynamics of model A and B, in the Hohenberg and Halperin classification [1], where a single field evolves towards equilibrium with non conserved order parameter dynamics (NCOP) or conserved order parameter dynamics (COP), respectively. On the contrary, the Phase field model [4,5] which describes the coupling of a NCOP system with a diffusive field such as temperature or concentration seems not to have been completely explored, in spite of the fact that it displays a variety of interesting features. For example, one could mention the formation of solid-liquid fronts during the solidification process, a phenomenon widespread in nature and the possibility of observing a vast class of patterns according to the initial conditions and external fields. An advancing solid front releases heat which in turn diffuses and promotes the solidification of the undercooled melt. One also observes that the shape of the moving front is subject to an instability due to the coupling with the diffusive field whose result is the formation of tips and dendritic growth. Such a phenomenon is due to the presence of the diffusive heat-mode at the interface which conspires to destabilize the solid-melt boundary as predicted by Mullins and Sekerka [6]. In the present Letter we study theoretically the evolution of an $N$-component version of the Phase field model and show that the coupling between the massless transverse modes and the diffusive field produce an instability which is the analogue of the Mullins-Sekerka instability.

We shall consider a suitable generalization of the Hamiltonian of Refs. [4,7,8] to two coupled $N$ component vector fields $\boldsymbol{\phi} = (\phi_1(\boldsymbol{x},t), ..., \phi_N(\boldsymbol{x},t))$ and $\boldsymbol{U} = (U_1(\boldsymbol{x},t), ..., U_N(\boldsymbol{x},t))$:

$$H[\boldsymbol{\phi}(\boldsymbol{x}),\ \boldsymbol{U}(\boldsymbol{x},t)] = \int d\boldsymbol{x} \Big[\frac{1}{2}(\nabla\boldsymbol{\phi})^2 + \frac{r}{2}\boldsymbol{\phi}^2 + \frac{g}{4N}(\boldsymbol{\phi}^2)^2 + \frac{w}{2}\boldsymbol{U}^2 + \mu \boldsymbol{U}\boldsymbol{\phi}\Big] \tag{1}$$

where $r$ and $g$ are the standard quadratic and quartic couplings of the Ginzburg-Landau model and the last term represent a bilinear coupling between the field $\boldsymbol{\phi}$ and $\boldsymbol{U}$. Stability imposes $g > 0$ and $w > 0$. As far as the equilibrium properties are involved, one sees that the $\boldsymbol{U}$ field can be traced out, and the Hamiltonian (1) reduces to the familiar $O(N)$ spherical model [9]. The evolution towards equilibrium is described by the two coupled Langevin equations:

$$\frac{\partial \phi_\alpha(\boldsymbol{x},t)}{\partial t} = -\Gamma_\phi \frac{\delta}{\delta \phi_\alpha(\boldsymbol{x},t)} H[\boldsymbol{\phi},\boldsymbol{U}] + \eta_\alpha(\boldsymbol{x},t) \tag{2}$$

$$\frac{\partial U_\alpha(\boldsymbol{x},t)}{\partial t} = \Gamma_U \nabla^2 \frac{\delta}{\delta U_\alpha(\boldsymbol{x},t)} H[\boldsymbol{\phi},\boldsymbol{U}] + \xi_\alpha(\boldsymbol{x},t) \tag{3}$$

Equation (2) represents the evolution of a NCOP $\boldsymbol{\phi}$ coupled to a COP $\boldsymbol{U}$, described by eq. (3). The noises appearing on the right hand sides of eqs. (2)-(3) have zero averages and are non conserved and conserved, respectively:

$$\langle \eta_\alpha(\boldsymbol{x},t)\eta_\beta(\boldsymbol{x}',t)\rangle = 2T_f\Gamma_\phi\delta_{\alpha,\beta}\delta(\boldsymbol{x}-\boldsymbol{x}')\delta(t-t') \tag{4}$$

$$\langle \xi_\alpha(\boldsymbol{x},t)\xi_\beta(\boldsymbol{x}',t)\rangle = -2T_f\Gamma_U\delta_{\alpha,\beta}\nabla^2\delta(\boldsymbol{x}-\boldsymbol{x}')\delta(t-t') \tag{5}$$

where $T_f$ is the temperature of the final equilibrium state whereas $\Gamma_U$ and $\Gamma_\phi$ are the kinetic coefficients. The noises $\xi_\alpha$ and $\eta_\alpha$ have vanishing mutual correlations.

To make contact with physically motivated models we notice that in the scalar case $N = 1$ eqs. (2)-(3) reduce to the well known Phase field model, devised by Langer [2,5], to understand the solidification process starting from an undercooled melt. To this purpose one performs the transformation $U = u - \mu\phi/w$ and identifies $\phi$ with the solid-liquid order parameter and $u$ with the temperature field shift from the melting value and obtains the familiar order parameter and heat diffusion equations.

In order to study the properties of the system it is sufficient to consider the three equal time real space connected correlation functions $C_{\phi\phi}(r,t) = \langle\phi_\alpha(R+r,t)\,\phi_\alpha(R,t)\rangle$, $C_{\phi U}(r,t) = \langle\phi_\alpha(R+r,t)\,U_\alpha(R,t)\rangle$ and $C_{UU}(r,t) = \langle U_\alpha(R+r,t)\,U_\alpha(R,t)\rangle$ which are independent of the index $\alpha$ by symmetry.

Within the $N\to\infty$ limit the dynamical equations for the correlations can be written in a closed form and the exact asymptotic behaviour explicitly extracted. In this limit their Fourier transforms, the structure functions, $\tilde{C}_{ij}(k,t)$, evolve according to the set of coupled equations:

$$\frac{1}{2}\frac{\partial\tilde{C}_{\phi\phi}(k,t)}{\partial t} = M_{\phi\phi}(k,t)\tilde{C}_{\phi\phi}(k,t) + M_{\phi U}(k,t)\tilde{C}_{\phi U}(k,t) + \Gamma_\phi T_f \qquad (6)$$

$$\frac{\partial\tilde{C}_{\phi U}(k,t)}{\partial t} = M_{U\phi}(k,t)\tilde{C}_{\phi\phi}(k,t) + (M_{UU}(k,t) + M_{\phi\phi}(k,t))\tilde{C}_{\phi U}(k,t) + M_{\phi U}(k,t)\tilde{C}_{UU}(k,t) \qquad (7)$$

$$\frac{1}{2}\frac{\partial\tilde{C}_{UU}(k,t)}{\partial t} = M_{U\phi}(k,t)\tilde{C}_{\phi U}(k,t) + M_{UU}(k,t)\tilde{C}_{UU}(k,t) + \Gamma_U T_f k^2. \qquad (8)$$

The matrix elements $M$ are given by

$$\begin{array}{ll} M_{\phi\phi}(k,t) = & -\Gamma_\phi(k^2+r+gS), \\ M_{\phi U}(k,t) = & -\Gamma_\phi\mu \\ M_{U\phi}(k,t) = & -\Gamma_U\mu k^2, \quad M_{UU}(k,t) = -\Gamma_U w k^2 \end{array} \qquad (9)$$

where $S(t) = \frac{1}{(2\pi)^d}\int_{|k|<\Lambda} d^dk\,\tilde{C}_{\phi\phi}(k,t)$ contains a phenomenological momentum cutoff and $d$ is the space dimensionality. The model for $r < \mu^2/w$, $g > 0$ and $d \geq 3$ displays a high temperature paramagnetic phase and a low temperature ordered phase, for $T_f < T_c$, the critical temperature of the model. As found in models A and B, $T_c = (\mu^2/w - r)(d-2)/(g\Lambda^{d-2}K_d)$ with $1/K_d = 2^{d-1}\pi^{d/2}\Gamma(d/2)$ where $\Gamma(x)$ is the usual Gamma function, and the structure factor reads

$$\lim_{t\to\infty}\tilde{C}_{\phi\phi}(k,t) = \frac{T_f}{k^2+r+gS(\infty)-\mu^2/w}. \qquad (10)$$

In other words at equilibrium the diffusive field produces a trivial renormalization of the bare parameter $r$. The quantity $(r+gS(t)-\mu^2/w)$ vanishes asymptotically and the small $k$ structure function diverges. Dynamically, instead, the presence of $\boldsymbol{U}$ induces non trivial effects on the field $\boldsymbol{\phi}$ because it acts on a time scale longer than the noise field, characterized by a short correlation time. In the following we shall treat the two fields on equal footing. The line $T_c(\mu)$ separates a high temperature phase from a low temperature phase characterized by infinite correlation length and divergent fluctuations at zero wavevector, due to the presence of massless Goldstone modes.

Since the behavior at $T_f = 0$ is representative of the entire ordered phase at $T_f < T_c$ we can neglect the noise term in the following analysis [10].

In the rest of the Letter we shall discuss the analytical predictions on the dynamical behaviour. These are compared with the numerical solution of eqs. (6)-(8) for quenches at $T_f = 0$ starting from uncorrelated random initial conditions, $d = 3$, $\Lambda = 1$ and $r < 0$.

In order to obtain a physical understanding of the relaxation process we notice that the mode coupling term $S(t)$ varies slowly with respect to the independent variable $t$. Thus a local linear analysis of the system of equations (2)-(3) for $T_f = 0$ is sufficient to extract the three kinds of regimes which characterize the approach to equilibrium. The linearized system has, for each value of $k$, two linearly independent solutions, proportional to $\exp[\omega_+(k)t]$ and $\exp[\omega_-(k)t]$, associated with the two eigenvalues $\omega_+(k)$ and $\omega_-(k)$ of the matrix $M$.

The larger eigenvalue $\omega_+(k)$ controls the dynamics and a small $k$ expansion is appropriate to extract the qualitative behaviour of the solutions. For temperatures below $T_c$ and in the early regime $S(t)$ is small. Thus, since $(r+gS(t)) < 0$ one finds:

$$\omega_+(k) = \Gamma_\phi|r+gS| - \left(\Gamma_\phi - \Gamma_U\frac{\mu^2}{|r+gS|}\right)k^2 + c_4k^4 \qquad (11)$$

where $c_4$ is a negative coefficient. One observes $\tilde{C}_{\phi\phi}(k,t)$ to display a peak around $k = 0$, growing in time with a power law $\sim t^{d/2}$, as in the pure NCOP, i.e. the longest wavelength fluctuations grow faster. This is due to the fact that $\tilde{C}_{\phi\phi}(k,t)$ grows exponentially with a rate approximately $-2\Gamma_\phi(r+gS+k^2)$. The coupling $\mu$ to the diffusive field has little effect in this regime and the system relaxes as if it had to reach an equilibrium characterized by $r+gS = 0$. However, this equilibrium is never achieved because of the presence of the field $\boldsymbol{U}$. In fact, after a characteristic time $\tau_f$ the value of $r+gS(t)$ vanishes due to the build-up of correlations and the dispersion relation becomes:

$$\omega_+(k) = \sqrt{\Gamma_\phi\Gamma_U}\mu k - \frac{\Gamma_\phi+\Gamma_U w}{2}k^2 + O(k^3). \qquad (12)$$

Within this second regime the system evolves as if $r = 0$ $g = 0$ and we observe the growth of correlations with finite wavevectors. This fact has remarkable repercussions on the structure factor, because we observe the appearance of a fastest growing mode with $k = k_f > 0$. At $t = \tau_f$, $k_f$ jumps discontinuously from 0 to a finite value. The position $k_m(t)$ of the peak of the structure factor $\tilde{C}_{\phi\phi}(k,t)$ then moves towards larger values of $k$. This regime describes formation of spatially periodic structures. This mechanism reminds the Mullins-Sekerka instability where the solid-melt interface becomes unstable to finite wavelength fluctuations and one observes pattern formation or dendrites.

Since $S(t)$ grows monotonically when $r+gS(t) > 0$ the dispersion relation takes the form:

$$\omega_+(k) = \Gamma_U \left( \frac{\mu^2}{r+gS} - w \right) k^2 + \tilde{c}_4 k^4 \qquad (13)$$

with $\tilde{c}_4 < 0$. Notice that the long wavelength modes become unstable because of the positive sign of the coefficient of order $k^2$. In fig. 1 we display the dispersion relation $\omega_+(k)$ for two typical times, $t < \tau_f$ and $t > \tau_f$. Here there exist values of $k$ for which $\omega_+(k)$ is positive, and a small perturbation with wavevectors falling within this range will be amplified. From the quasilinear analysis one can extract the value at which the fastest growing mode appears for the first time. For small $\mu$ this occurs at the wavevector:

$$k_f = \mu \frac{\sqrt{\Gamma_\phi \Gamma_U}}{\Gamma_\phi + \Gamma_U w}. \qquad (14)$$

From scaling we argue that the criterion for the instability to occur is that the domain size $L(t) \sim t^{1/2}$ must exceed the characteristic length $\lambda_f$ given by $2\pi k_f^{-1}$, therefore $\tau_f \sim \mu^{-2}$ for small $\mu$. In fig. 2 we display the numerical results for $\tau_f$ vs. $\mu^{-2}$ for small $\mu$, the agreement is remarkable. Beyond the linear analysis one finds that the quartic coupling hinders the unbounded growth of fluctuations. The non linear term, in fact, has a twofold role since initially it promotes the formation of the peak of the structure factor at finite wavevectors, but finally prevents it from growing indefinitely and eventually drives it towards zero wavevectors.

At very late times, in the regime of the third kind, the peak finally tends towards $k = 0$ and the width shrinks as the system tends to equilibrate. As $t \to \infty$ and $r + gS - \mu^2/w \to 0^-$, the coefficient of order $k^2$ in eq. (13) vanishes and the system tends towards the COP static fixed point, displaying the typical behavior of conserved dynamics. Within this regime the domain size grows as $L(t) \sim t^{1/4}$ and $k_m(t) \sim (\ln(t)/t)^{1/4}$ as predicted by the exact solution of the corresponding COP model without diffusion ($\mu = 0$). These predictions agree with the numerical solution. In fig. 3 the structure factors at different times are shown.

These findings are consistent with the existence of multiscaling in the very late regime ($t \gg \tau_f$), a particular form of scaling where the exponent is scale dependent. In particular the dynamical structure factor clearly displays a multiscaling behavior as shown in $\tilde{C}_{\phi\phi}(k,t) \sim [L(t)^2 k_m(t)^{2-d}]^{\varphi(k/k_m)}$ [9] as shown in fig. 4. Such a multiscaling behavior follows from the competition of two marginally distinct lengths, namely the domain size $L(t)$ and $k_f^{-1}$.

To conclude we have studied a model which provides an interesting new scenario. The observed instability at finite wavevectors is due to the coupling between low energy Nambu-Goldstone modes and the diffusive modes. We believe this is a paradigm for the Mullins-Sekerka type of phenomena where the soft modes are represented by the capillary wave spectrum associated with the solid-melt interface and the diffusive mode is the heat transport. These two fields concur to destabilize the solid-melt boundary in analogy with our findings. On physical grounds, one expects this kind of instability to occur during phase separation, because small droplets can dissipate heat more efficiently and reach rapidly thermal equilibrium due to the Gibbs-Thomson effect, whereas larger droplets try to dissipate energy faster by creating bulges thus increasing the curvature. As the system cools down upon reaching equilibrium the typical length of the bulges, $\lambda_f$, increases and diverges together with the average domain-size $L(t)$.

Our study of the present model is non perturbative and reveals an unexpectedly rich dynamical behavior: it displays scaling behavior in the early regime ($t < \tau_f$) and multiscaling [9] in the late regime ($t \gg \tau_f$). Moreover it displays a transient characterized by an instability which reminds in many respects the one occuring in solidification kinetics. It remains to be explored in greater detail the role of fluctuations although the noise is thought to be irrelevant below the critical temperature of the model.
We thank M. Conti and L. Peliti for illuminating discussions. This work was partially supported by grants from INFM and INFN.

---

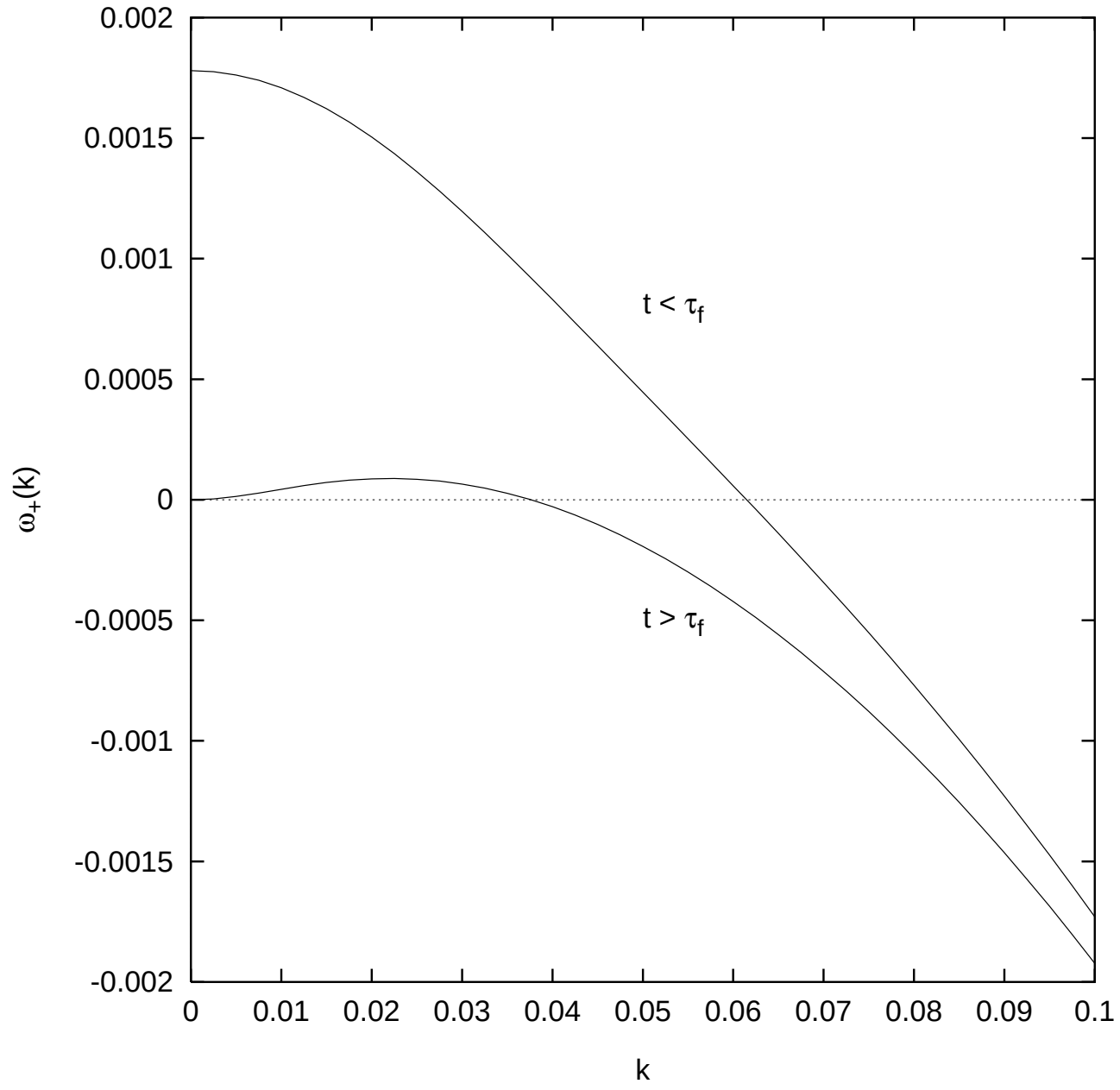

FIG. 1. Dispersion relation $\omega_+(k)$ as a function of $k$ for different regimes: early regime (upper curve), and late regime (lower curve).

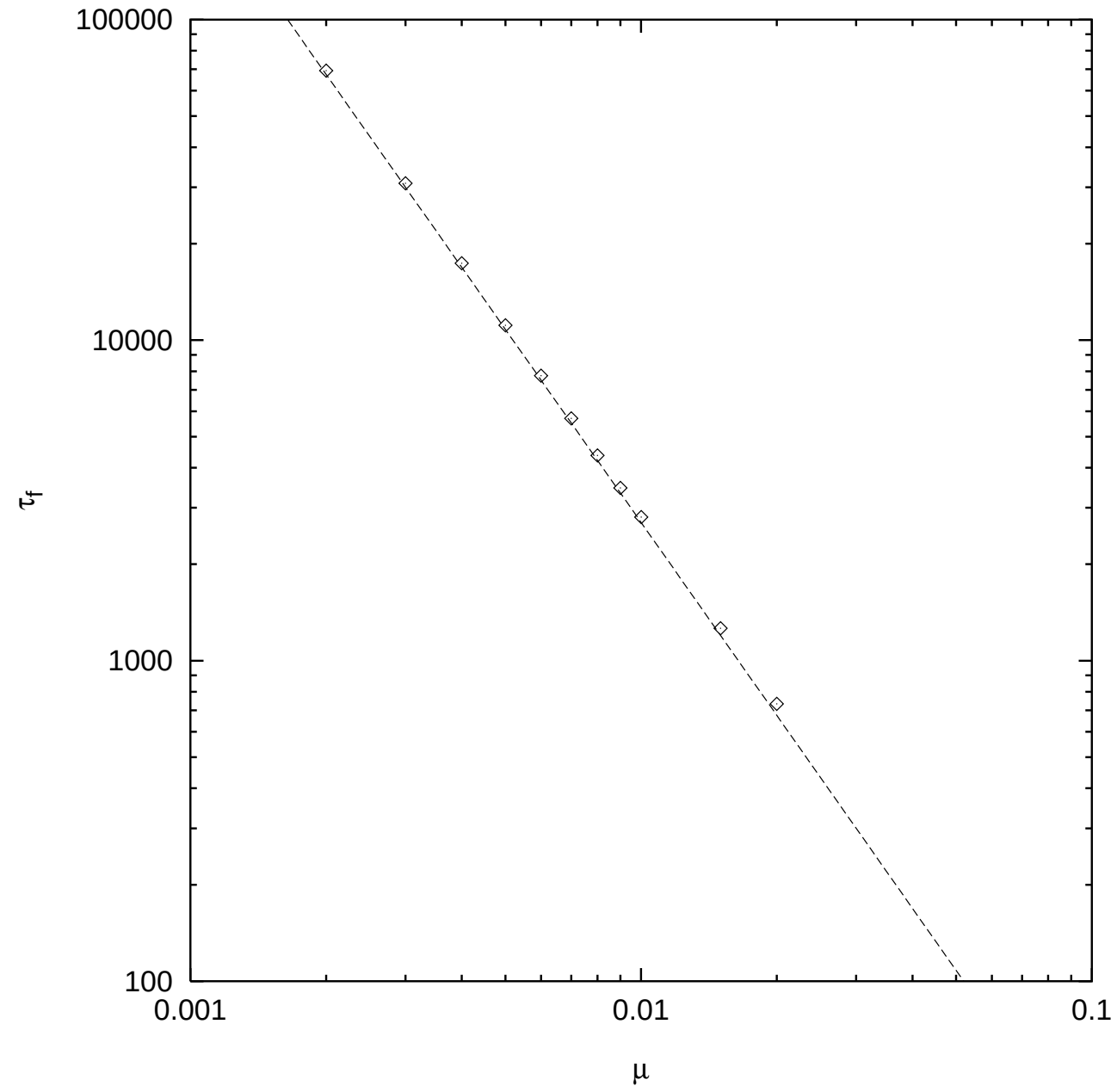

FIG. 2. Scaling of $\tau_f$ as function of $\mu$. The broken line has slope $-2$. The squares are obtained from the numerical solution with $\Gamma_\phi = 1$, $\Gamma_U = 5$, $r = -0.5$, $g = 1$, $w = 0.05$ and $d = 3$.

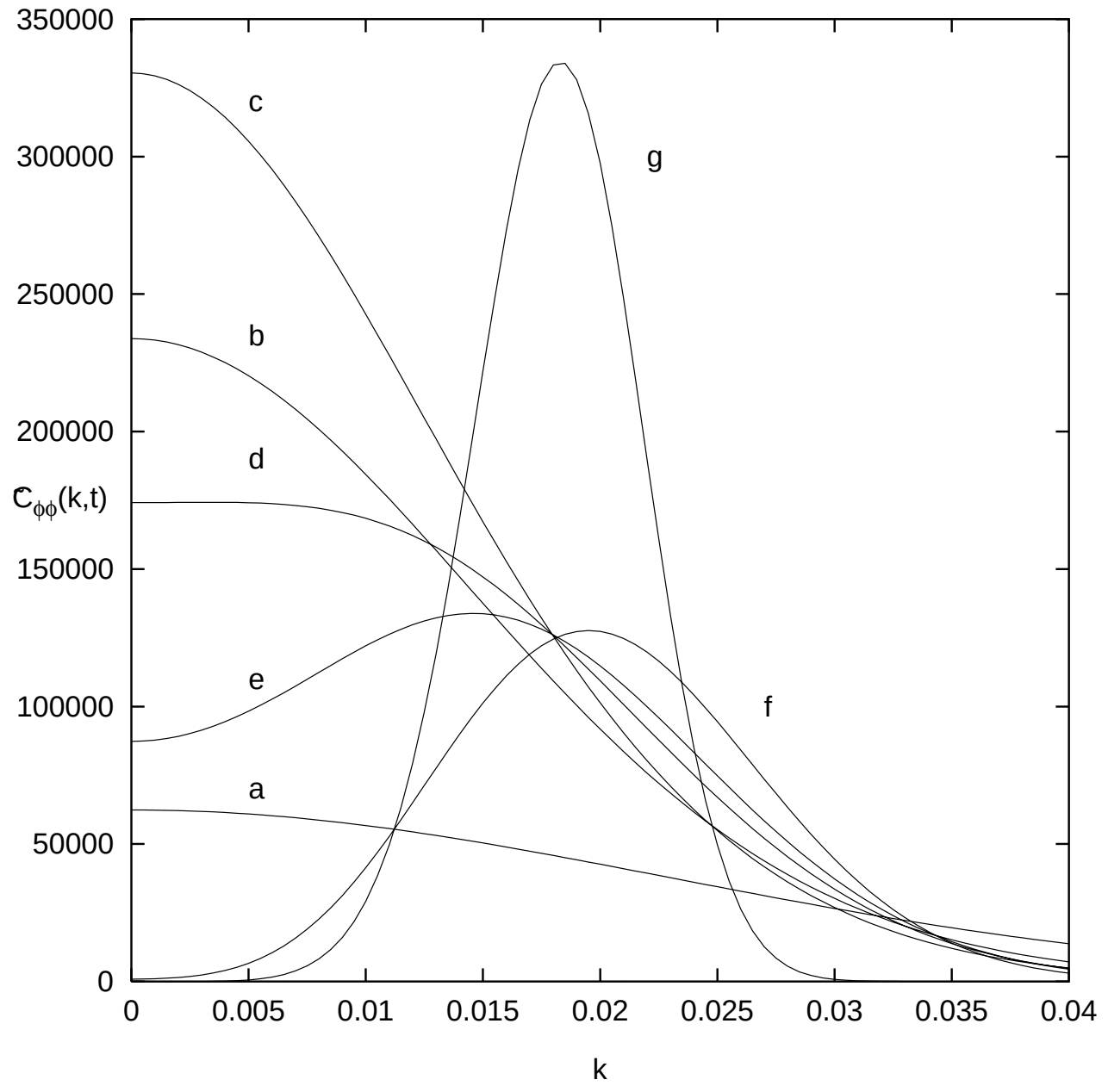

FIG. 3. Structure factor $\tilde{C}_{\phi\phi}(k,t)$ for $\mu = 0.01$ at successive times: a) $t = 500$, b) $t = 1500$, c) $t = 3000$, d) $t = 5000$, e) $t = 6000$, f) $t = 9000$, g) $t = 80000$. The parameters are the same as fig. 2.

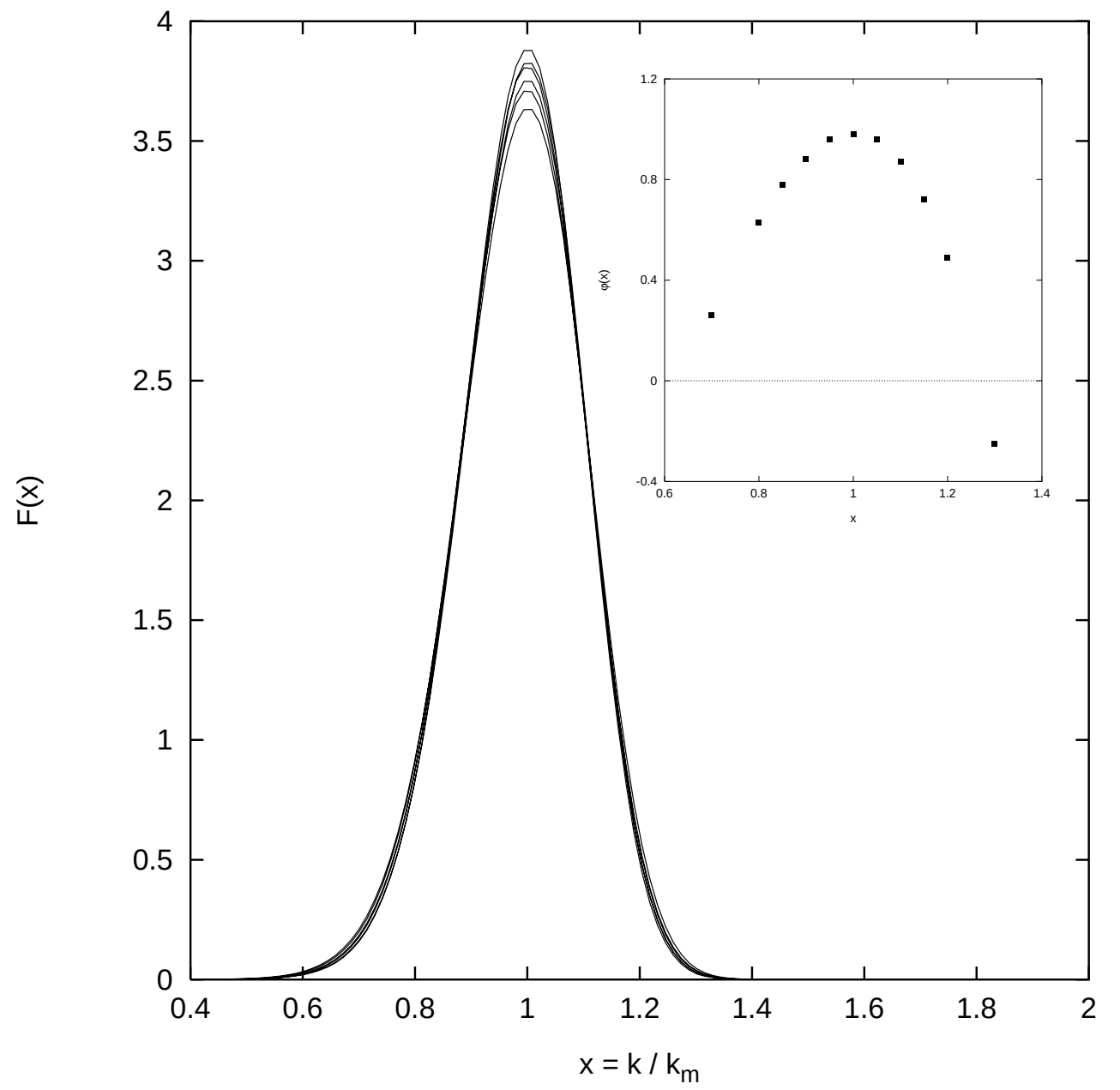

FIG. 4. Shape function $F(x) = k_m^d(t)\ \tilde{C}_{\phi\phi}(xk_m, t)$ as a function of $x = k/k_m$ for different times for $\mu = 0.01$. The non perfect data collapse signals multiscaling. In the inset we report the numerical value of the multiscaling exponent $\varphi(x)$ defined in the text obtained from the best fit of $C_{\phi\phi}(xk_m, t)$ as a function of $L(t)^2 k_m(t)^{2-d}$. The parameters are the same as fig. 2.